\documentstyle[12pt]{article}

\begin{document}

\author{Yishi Duan and Ying Jiang\thanks{Corresponding author; E--mail: itp3@lzu.edu.cn}\\
{\small \it Institute of Theoretical Physics, Lanzhou University, Lanzhou
730000, P. R. China}\\
\\
Guohong Yang\\
{\small \it Department of Physics, Fudan University, Shanghai 200433, P. R. China}}

\title{The topological quantization and bifurcation of the topological linear defects%
\thanks{Work supported by the National Natural Science Foundation of P. R. China}
}

\date{}
\maketitle

\begin{abstract}
\baselineskip18pt
In the light of $\phi $--mapping
method and topological current theory, the topological structure and the
topological quantization of topological linear defects are obtained
under the condition that the Jacobian $J(\frac \phi v)\neq 0$.
When $J(\frac \phi v)=0$, it is shown
that there exist the crucial case of branch process. Based on the implicit
function theorem and the Taylor expansion, the origin and bifurcation of the
linear defects are detailed in the neighborhoods of the limit points and
bifurcation points of $\phi $-mapping, respectively.

\end{abstract}

\vskip1.5cm

\baselineskip20pt

The world of topological linear defects is amazingly rich and
have been recently the focus of much attention in many areas of contemporary
physics\cite{sf,sc,14,3}. However, since most of them are based on some concrete models\cite
{9}, the discussions of the topological property of the linear defects are
not too clear. In this letter, we will discuss the topological structure,
topological quantization and the bifurcation of these linear defects through
the $\phi $--mapping and topological current theory\cite{11} in
four--dimensional curved space--time without regard to any concrete model.

As is well known, in general, the topological linear defects are determined 
by a 2--dimensional vector field $\vec \phi =(\phi ^1,\phi ^2)$ which is
just the ordered parameter\cite{14} or the condensate wave function 
(generally, a complex wave function $\Phi $ can be expressed by 
$\Phi =\phi ^1+i\phi ^2$, it is equivalent to a 2--dimensional vector field
$\vec \phi =(\phi ^1,\phi ^2)$)\cite{sf,sc,bec}, and it can be looked upon
as a smooth map $\phi:X\rightarrow
R^2$
\begin{equation}
\phi ^a=\phi ^a(x),\;\;\;\;a=1,2. 
\end{equation}
where $X$ be a 4-dimensional Riemannian manifold with metric tensor $g_{\mu
\nu }$ and local coordinates $x^\mu $ $(\mu ,\nu =1,...,4)$, and $R^2$ a
2--dimensional Euclidean space. The direction field of $\vec \phi (x)$ is 
generally determined by
\begin{equation}
\label{0}n^a=\frac{\phi ^a}{||\phi (x)||},\;\;\;\;||\phi (x)||=\sqrt{\phi
^a(x)\phi ^a(x)} 
\end{equation}
with 
\begin{equation}
\label{1}n^a(x)n^a(x)=1. 
\end{equation}
It is obvious that $n^a(x)$ is a section of the sphere bundle 
$S(X)$\cite{11}. If $n^a(x)$ is a smooth unit vector field without 
singularities or it has singularities somewhere but at the point 
$\vec \phi (x)\neq 0$, from (\ref{1}) we have 
\begin{equation}
\label{2}n^a\partial _\mu n^a=0,\;\;\;\mu =1,2,3,4, 
\end{equation}
which can be looked upon as a system of 4 homogenous linear equations of $%
n^a $ with coefficient matrix $[\partial _\mu n^a].$ The necessary and
sufficient condition that (\ref{2}) has non--trivial solution for $n^a(x)$
is $rank[\partial _\mu n^a]<2$, i.e. the Jacobian determinants 
\begin{equation}
\label{3}D^{\mu \nu }(\partial n)=\frac 12\epsilon ^{\mu \nu \lambda \rho
}\epsilon _{ab}\partial _\lambda n^a\partial _\rho n^b 
\end{equation}
are equal to zero. However, at the point $\vec \phi (x)=0$, the above
consequences are not held. In fact, $D^{\mu \nu }(\partial n)$ behave 
themselves like a $\delta$--function $\delta(\vec \phi )$  
\begin{equation}
\label{4}D^{\mu \nu }(\partial n)=\left\{ 
\begin{array}{ccc}
=0, & \;for\; & \vec \phi (x)\neq 0, \\ 
\neq 0, & \;for\; & \vec \phi (x)=0, 
\end{array}
\right. 
\end{equation}
which will be shown in the following. So we are focussed on the zeroes of $\phi ^a(x).$

Suppose that for the system of equations $\vec \phi(x)=0$, there are $l$ different solutions, when the solutions are regular at which
the rank of the Jacobian matrix $[\partial _\mu \phi ^a]$ is 2, the
solutions of $\vec \phi (x)=0$ can be expressed parameterizedly by 
\begin{equation}
\label{5}x^\mu =z_i^\mu (u^1,u^2),\,\,\,i=1,\cdot \cdot \cdot l,
\end{equation}
where the subscript $i$ represents the $i$-th solution and the parameters $%
u^I(I=1,2)$ span a 2-dimensional submanifold with the metric tensor $%
g_{IJ}=g_{\mu \nu }\frac{\partial x^\mu }{\partial u^I}\frac{\partial x^\nu 
}{\partial u^J}$ which is called the $i$-th singular submanifold $N_i$ of $%
\phi $--mapping in $X$. For each $N_i$, we can define a normal submanifold $%
M_i$ in $X$ which is spanned by the parameters $v^A(A=1,2)$ with the metric
tensor $g_{AB}=g_{\mu \nu }\frac{\partial x^\mu }{\partial v^A}\frac{%
\partial x^\nu }{\partial v^B}$, and the intersection point of $M_i$ and $N_i
$ is denoted by $p_i$which can be expressed parameterizedly by $v^A=p_i^A$.
By virtue of the implicit function theorem, it should be hold true that, 
at the regular point $p_i$, the Jacobian matrix $J(\frac \phi v)$ satisfies 
\begin{equation}
\label{nonzero}J(\frac \phi v)=\frac{D(\phi ^1,\phi ^2)}{D(v^1,v^2)}\neq 0.
\end{equation}

In the following, we will induce a rank--two topological tensor current
through the integration of $D^{\mu \nu }(\partial n)$ in (\ref{3}) on $M_i$.
As is well known, the Winding number $W_i$ of $\vec \phi (x)$ on $M_i$ at $%
p_i$ is expressed by\cite{19}
\begin{equation}
\label{6}W_i=\frac 1{2\pi }\int_{\partial \Sigma _i}n^{*}(\epsilon
_{ab}n^adn^b), 
\end{equation}
where $\partial \Sigma _i$ is the boundary of a neighborhood $\Sigma _i$ of
$p_i$ on the two--surface $M_i$ with $p_i\notin \partial \Sigma _i$ and
$\Sigma _i\cap \Sigma _j=\emptyset$, $n^{*}$ is the pull back of the Gauss map
$n:\partial \Sigma_i \rightarrow S^1$.  It is
known that the Winding
numbers $W_i$ are corresponding to the first homotopy group $\pi [S^1]=Z$ (
the set of integers). In topology, it means that, when the point $x^\mu $ or $v^A$ covers $\partial
\Sigma _i$ once, the unit vector $n^a$ will cover $S^1$ $W_i$ times, which
is a topological invariant and is also called the degree of Gauss map. Using
the Stokes' theorem in the exterior differential form, one can deduced that%
\begin{eqnarray}
W_i&=&\frac 1{2\pi }\int_{\partial \Sigma _i}\epsilon _{ab}n^a\partial _\rho
n^bdx^\rho \nonumber\\
\label{7}&=&\frac 1{2\pi }\int_{\Sigma _i}\frac 1{\sqrt{g_x}}D^{\mu \nu
}(\partial n)d\sigma _{\mu \nu } 
\end{eqnarray}
where $d\sigma _{\mu \nu }$ is the invariant surface element of $\Sigma _i$
and $g_x=\det (g_{\mu \nu }).$ As mentioned above, the deduction (\ref{7})
shows that the Winding number $W_i$ can be expressed as the integration of $%
D^{\mu \nu }(\partial n)$ on $M_i.$

From above discussion, especially the expressions (\ref{3}), (\ref{4}) and (%
\ref{7}), we can induce a rank--two topological tensor current $j^{\mu \nu }$%
corresponding to the Winding number, 
\begin{equation}
\label{8}j^{\mu \nu }=\frac 1{2\pi }\frac 1{\sqrt{g_x}}\epsilon ^{\mu \nu
\lambda \rho }\epsilon _{ab}\partial _\lambda n^a\partial _\rho n^b.
\end{equation}
Obviously, $j^{\mu \nu }$ is antisymmetric and identically conserved. By the 
use of (\ref{3}) and (\ref{8}), it is shown that the rank--two topological
tensor current $j^{\mu \nu}$ has the same property with $D^{\mu \nu}$, i.e.
$j^{\mu \nu}$ does not vanish only at where $\vec \phi (x)=0$. 

From (\ref{4}) and (\ref{8}), we see that the core of the topological tensor
current $j^{\mu \nu }$ is the Kernel of $\phi $--mapping in $X$. We think
this is the essence of topological current theory and $\phi $--mapping is
the key to study this topological theory of linear defects. By the use of (%
\ref{0}) and 
\begin{equation}
\partial _\mu n^a=\frac{\partial _\mu \phi ^a}{||\phi ||}+\phi ^a\partial
_\mu (\frac 1{||\phi ||}),\;\;\;\frac \partial {\partial \phi ^a}\ln ||\phi
||=\frac{\phi ^a}{||\phi ||^2}, 
\end{equation}
which should be looked upon as generalized functions, $j^{\mu \nu }$ can be
expressed by 
\begin{equation}
j^{\mu \nu }=\frac 1{2\pi }\frac 1{\sqrt{g_x}}\epsilon ^{\mu \nu \lambda
\rho }\epsilon _{ab}\frac \partial {\partial \phi ^c}\frac \partial
{\partial \phi ^a}\ln ||\phi ||\partial _\lambda \phi ^c\partial _\rho \phi
^b. 
\end{equation}
By defining the Jacobian determinants $J^{\mu \nu }(\frac \phi x)$ as 
\begin{equation}
\label{9}\epsilon ^{ab}J^{\mu \nu }(\frac \phi x)=\epsilon ^{\mu \nu \lambda
\rho }\partial _\lambda \phi ^a\partial _\rho \phi ^b 
\end{equation}
and making use of the 2--dimensional Laplacian Green function relation 
\begin{equation}
\Delta _\phi \ln ||\phi ||=2\pi \delta (\vec \phi ), 
\end{equation}
where $\Delta _\phi =(\frac{\partial ^2}{\partial \phi ^a\partial \phi ^a})$
is the 2--dimensional Laplacian operator in $\phi $--space, we do obtain the 
$\delta $--function structure of the topological tensor current rigorously 
\begin{equation}
\label{10}j^{\mu \nu }=\frac 1{\sqrt{g_x}}\delta (\vec \phi )J^{\mu \nu
}(\frac \phi x). 
\end{equation}
It is obvious that $j^{\mu \nu }$ is non--zero only when $\vec \phi =0$,
which is as expected.

In the following, we will detail the inner structure of $j^{\mu \nu }$ and
discuss the generation of linear defects and their topological quantization
from the rank--two tensor topological current in $X$.

As is well known\cite{17}, the definition of the $\delta $--function $\delta
(N_i)$ in curved space-time on a submanifold $N_i$ is
\begin{equation}
\label{m}\delta (N_i)=\int_{N_i}\frac 1{\sqrt{g_x}}\delta ^4(\vec x-\vec
z_i(u^1,u^2))\sqrt{g_u}d^2u. 
\end{equation}
where $g_u = \det(g_{IJ})$. Following this, by analogy with the procedure of deducing $\delta (f(x))$,
we can expand the $\delta $--function $\delta (\vec \phi )$ as 
\begin{equation}
\label{delta}\delta (\vec \phi )=\sum_{i=1}^lc_i\delta (N_i), 
\end{equation}
where the coefficients $c_i$ must be positive, i.e. $c_i=\mid c_i\mid $.
From the definition of $W_i$ in (\ref{6}), the Winding number can also be
rewritten in terms of the parameters $v^A$ of $M_i$ as 
\begin{eqnarray}
W_i&=&\frac 1{2\pi }\int_{\Sigma _i}\epsilon _{ab}\partial _An^a\partial
_Bn^bdv^A\wedge dv^B \nonumber\\
&=&\frac 1{2\pi }\int_{\Sigma _i}\epsilon ^{AB}\epsilon
_{ab}\partial _An^a\partial _Bn^bd^2v,
\end{eqnarray}
Then, by duplicating the above process, this integral leads to the result that
\begin{equation}
c_i=\frac{\beta _i\sqrt{g_v}}{\mid J(\frac \phi v)_{p_i}\mid }=\frac{\beta
_i\eta _i\sqrt{g_v}}{J(\frac \phi v)_{p_i}},\,\,\,\,g_v =\det(g_{AB}),
\end{equation}
where $\beta _i=|W_i|$ is a positive integer called the Hopf index\cite{20}
of $\phi $-mapping on $M_i,$ it means that when the point $v$ covers the
neighborhood of the zero point $p_i$ once, the function $\vec \phi $ covers
the corresponding region in $\vec \phi $-space $\beta _i$ times, and $\eta
_i=signJ(\frac \phi v)_{p_i}=\pm 1$ is the Brouwer degree of $\phi $-mapping%
\cite{20}. Substituting this expression of $c_i$ and (\ref{delta}) into (\ref
{10}), we gain the total expansion of the topological tensor current
\[
j^{\mu \nu }=\frac 1{\sqrt{g_x}}\sum_{i=1}^l\frac{\beta _i\eta _i\sqrt{g_v}}{%
J(\frac \phi v)|_{p_i}}\delta (N_i)J^{\mu \nu }(\frac \phi x) 
\]
or in terms of parameters $y^A=(v^1,v^2,u^1,u^2)$%
\begin{equation}
j^{AB}=\frac{1}{\sqrt{g_y}}\sum_{i=1}^l\frac{\beta _i\eta _i\sqrt{g_v}}{%
J(\frac \phi v)|_{p_i}}\delta (N_i)J^{AB}(\frac \phi y),\,\,\,A,B=1,2,3,4.
\end{equation}
From the above equation, we conclude that the inner structure of $j^{\mu \nu
}$ or $j^{AB}$ is labelled by the total expansion of $\delta (\vec \phi )$,
which includes the topological information $\beta _i$ and $\eta _i.$

It is obvious that, in (\ref{5}), when $u^1$ and$\,u^2$ are taken to be
time-like evolution parameter and space-like string parameter respectively,
the inner structure of $j^{\mu \nu }$ just represents $l$ topological linear
defects moving in $X$, the 2-dimensional singular submanifolds $%
N_i\,\,(i=1,\cdot \cdot \cdot l)$ are their world sheets. Here we see that
the topological defects are generated from where $\vec \phi=0$ and does not
tie on any concrete models. Furthermore, we see that the Hopf indices $\beta
_i$ and Brouwer degrees $\eta _i$ classify these linear defects. In detail,
the Hopf indices $\beta _i$ characterize the absolute values of the
topological quantization and the Brouwer degrees $\eta _i=+1$ correspond to
defects while $\eta _i=-1$ to antidefects.

Following, in order to discuss the bifurcation property of these topological
linear defects and to simplify our study, we select the parameter $u^1$ as
the evolution parameter $t$, and let the string parameter $u^2=\sigma $ be
fixed. In this case, the Jacobian matrices are reduced to%
\[
J^{A4}\equiv J^A,\;\;\;J^{AB}=0,\;\;J^3=J^{34}=J(\frac \phi
v),\;\;\;\;A,B=1,2,3, 
\]
for $y^4=u^2\equiv \sigma $. In the above we have studied the
topological property of the multidefects in the
case that the vector field $\vec \phi $ only consists of regular points,
i.e. (\ref{nonzero}) is hold true. In the following, we will study the case
when (\ref{nonzero}) fails. It often happens when the zeros of $\vec \phi $
include some branch points, which lead to the bifurcation of the
topological current. The branch points are determined by
\begin{equation}
\label{88}\left\{ 
\begin{array}{l}
\phi ^1(v^1,v^2,t,\sigma )=0 \\ 
\phi ^2(v^1,v^2,t,\sigma )=0 \\ 
\phi ^3(v^1,v^2,t,\sigma )\equiv J(\frac \phi v)=0 
\end{array}
\right. 
\end{equation}
for the fixed $\sigma $, and they are denoted as $(t^{*},p_i)$. In $\phi $%
-mapping theory usually there are two kinds of branch points, namely the
limit points and the bifurcation points\cite{27}, satisfying 
\begin{equation}
\label{89}J^A(\frac \phi y)|_{(t^{*},p_i)}\neq 0,\;\;\;\;\;A=1,2 
\end{equation}
and 
\begin{equation}
\label{90}J^A(\frac \phi y)|_{(t^{*},p_i)}=0,\;\;\;\;\;A=1,2 
\end{equation}
respectively. (For simplicity, we consider $A=1$ only.) First, we
will study the case (\ref{89}).

For the purpose of using the implicit function theorem to study the branch
properties of topological linear defects at the limit points, we use the Jacobian $J^1(\frac \phi
y)$ instead of $J(\frac \phi v)$ to search for the solutions of $\vec \phi
=0 $. This means we have replaced $v^1$ by $t$. For clarity we rewrite the
first two equations of (\ref{88}) as 
\begin{equation}
\label{91}\phi ^a(t,v^2,v^1,\sigma )=0,\ \;\;\;\;a=1,2. 
\end{equation}
Taking account of (\ref{89}) and using the implicit function theorem, we
have a unique solution of the equations (\ref{91}) in the neighborhood of
the limit point $(t^{*},p_i)$ 
\begin{equation}
\label{92}t=t(v^1,\sigma ),\ \;\;\;\;v^2=v^2(v^1,\sigma ) 
\end{equation}
with $t^{*}=t(p_i^1,\sigma )$. In order to show the behavior of the linear
defects at
the limit points, we will investigate the Taylor expansion of (\ref{92}) in
the neighborhood of $(t^{*},p_i)$. In the present case, from (\ref{89}) and
the last equation of (\ref{88}), we get 
\[
\frac{dv^1}{dt}=\frac{J^1(\frac \phi y)}{J(\frac \phi v)}|_{(t^{*},p_i)}=%
\infty 
\]
i.e. 
\[
\frac{dt}{dv^1}|_{(t^{*},p_i)}=0. 
\]
Then, the Taylor expansion of $t=t(v^1,\sigma )$ at the limit point $%
(t^{*},p_i)$ leads to
\begin{equation}
\label{93}t-t^{*}=\frac 12\frac{d^2t}{(dv^1)^2}|_{(t^{*},p_i)}(v^1-p_i^1)^2 
\end{equation}
which is a parabola in $v^1$---$t$ plane. From (\ref{93}) we can obtain two
solutions $v_{(1)}^1(t,\sigma )$ and $v_{(2)}^1(t,\sigma )$, which give the
branch solutions of the linear defects at the limit points. If $\frac{d^2t}{%
(dv^1)^2}|_{(t^{*},z_i)}>0$, we have the branch solutions for $t>t^{*}$,
otherwise, we have the branch solutions for $t<t^{*}$. The former is related
to the origin of linear defects at the limit points. Since the topological
current of multidefects is identically conserved, the topological quantum
numbers of these two generated linear defects must be opposite at the limit
point, i.e. $\beta _1\eta _1+\beta _2\eta _2=0$, which is important in the
early universe because of spontaneous symmetry
breaking\cite{4}.

Following, let us turn to consider the other case (\ref{90}), in
which the restrictions are 
\begin{equation}
\label{94}J(\frac \phi v)|_{(t^{*},p_i)}=0,\ \;\;\;\;J^1(\frac \phi
y)|_{(t^{*},p_i)}=0. 
\end{equation}
These two restrictive conditions will lead to an important fact that the
function relationship between $t$ and $v^1$ is not unique in the
neighborhood of bifurcation point $(t^{*},p_i)$. The equation 
\begin{equation}
\label{95}\frac{dv^1}{dt}=\frac{J^1(\frac \phi y)}{J(\frac \phi v)}%
|_{(t^{*},p_i)} 
\end{equation}
which under restraint of (\ref{94}) directly shows that the direction of the
integral curve of (\ref{95}) is indefinite at the point $(t^{*},p_i)$. This
is why the very point $(t^{*},p_i)$ is called a bifurcation point of the
multidefects current. With the aim of finding the different directions of
all branch curves at the bifurcation point, we suppose that 
\begin{equation}
\label{96}\frac{\partial \phi ^1}{\partial v^2}|_{(t^{*},p_i)}\neq 0. 
\end{equation}
From $\phi ^1(v^1,v^2,t,\sigma )=0$, the implicit function theorem says that
there exists one and only one function relationship 
\begin{equation}
\label{97}v^2=v^2(v^1,t,\sigma ) 
\end{equation}
with the partial derivatives $f_1^2=\partial v^2/\partial v^1$, $%
f_t^2=\partial v^2/\partial t$. Substituting (\ref{97}) into $\phi ^1$, we
have 
\[
\phi ^1(v^1,u^2(v^1,t,\sigma ),t,\sigma )\equiv 0 
\]
which gives 
\begin{equation}
\label{98}\frac{\partial \phi ^1}{\partial v^2}f_1^2=-\frac{\partial \phi ^1%
}{\partial v^1},\ \;\;\;\;\;\;\frac{\partial \phi ^1}{\partial v^2}f_t^2=-%
\frac{\partial \phi ^1}{\partial t}, 
\end{equation}
\[
\frac{\partial \phi ^1}{\partial v^2}f_{11}^2=-2\frac{\partial ^2\phi ^1}{%
\partial v^2\partial v^1}f_1^2-\frac{\partial ^2\phi ^1}{(\partial v^2)^2}%
(f_1^2)^2-\frac{\partial ^2\phi ^1}{(\partial v^1)^2}, 
\]
\[
\frac{\partial \phi ^1}{\partial v^2}f_{1t}^2=-\frac{\partial ^2\phi ^1}{%
\partial v^2\partial t}f_1^2-\frac{\partial ^2\phi ^1}{\partial v^2\partial
v^1}f_t^2-\frac{\partial ^2\phi ^1}{(\partial v^2)^2}f_t^2f_1^2-\frac{%
\partial ^2\phi ^1}{\partial v^1\partial t}, 
\]
\[
\frac{\partial \phi ^1}{\partial v^2}f_{tt}^2=-2\frac{\partial ^2\phi ^1}{%
\partial v^2\partial t}f_t^2-\frac{\partial ^2\phi ^1}{(\partial v^2)^2}%
(f_t^2)^2-\frac{\partial ^2\phi ^1}{\partial t^2}, 
\]
where 
\[
f_{11}^2=\frac{\partial ^2v^2}{(\partial v^1)^2},\ \;\;\;f_{1t}^2=\frac{%
\partial ^2v^2}{\partial v^1\partial t},\ \;\;\;f_{tt}^2=\frac{\partial ^2v^2%
}{\partial t^2}. 
\]
From these expressions we can calculate the values of $%
f_1^2,f_t^2,f_{11}^2,f_{1t}^2$ and $f_{tt}^2$ at $(t^{*},p_i)$.

In order to explore the behavior of the topological linear defects at the
bifurcation points, let us investigate the Taylor expansion of 
\begin{equation}
\label{99}F(v^1,t,\sigma )=\phi ^2(v^1,v^2(v^1,t,\sigma ),t,\sigma ) 
\end{equation}
in the neighborhood of $(t^{*},p_i)$, which according to the Eqs.(\ref{88})
must vanish at the bifurcation point, i.e. 
\begin{equation}
\label{100}F(t^{*},p_i)=0. 
\end{equation}
From (\ref{99}), the first order partial derivatives of $F(v^1,t,\sigma )$
with respect to $v^1$ and $t$ can be expressed by 
\begin{equation}
\label{101}\frac{\partial F}{\partial v^1}=\frac{\partial \phi ^2}{\partial
v^1}+\frac{\partial \phi ^2}{\partial v^2}f_1^2,\ \;\;\;\frac{\partial F}{%
\partial t}=\frac{\partial \phi ^2}{\partial t}+\frac{\partial \phi ^2}{%
\partial v^2}f_t^2. 
\end{equation}
Making use of (\ref{98}), (\ref{101}) and Cramer's rule, it is easy to prove
that the two restrictive conditions (\ref{94}) can be rewritten as 
\[
J(\frac \phi v)|_{(t^{*},p_i)}=(\frac{\partial F}{\partial v^1}\frac{%
\partial \phi ^1}{\partial v^2})|_{(t^{*},p_i)}=0, 
\]
\[
J^1(\frac \phi y)|_{(t^{*},p_i)}=(\frac{\partial F}{\partial t}\frac{%
\partial \phi ^1}{\partial v^2})|_{(t^{*},p_i)}=0, 
\]
which give 
\begin{equation}
\label{102}\frac{\partial F}{\partial v^1}|_{(t^{*},p_i)}=0,\ \;\;\;\frac{%
\partial F}{\partial t}|_{(t^{*},p_i)}=0 
\end{equation}
by considering (\ref{96}). The second order partial derivatives of the
function $F$ are easily to find out to be 
\[
\frac{\partial ^2F}{(\partial v^1)^2}=\frac{\partial ^2\phi ^2}{(\partial
v^1)^2}+2\frac{\partial ^2\phi ^2}{\partial v^2\partial v^1}f_1^2+\frac{%
\partial \phi ^2}{\partial v^2}f_{11}^2+\frac{\partial ^2\phi ^2}{(\partial
v^2)^2}(f_1^2)^2 
\]
\[
\frac{\partial ^2F}{\partial v^1\partial t}=\frac{\partial ^2\phi ^2}{%
\partial v^1\partial t}+\frac{\partial ^2\phi ^2}{\partial v^2\partial v^1}%
f_t^2+\frac{\partial ^2\phi ^2}{\partial v^2\partial t}f_1^2+\frac{\partial
\phi ^2}{\partial v^2}f_{1t}^2+\frac{\partial ^2\phi ^2}{(\partial v^2)^2}%
f_1^2f_t^2 
\]
\[
\frac{\partial ^2F}{\partial t^2}=\frac{\partial ^2\phi ^2}{\partial t^2}+2%
\frac{\partial ^2\phi ^2}{\partial v^2\partial t}f_t^2+\frac{\partial \phi ^2%
}{\partial v^2}f_{tt}^2+\frac{\partial ^2\phi ^2}{(\partial v^2)^2}(f_t^2)^2 
\]
which at $(t^{*},p_i)$ are denoted by 
\begin{equation}
\label{103}A=\frac{\partial ^2F}{(\partial v^1)^2}|_{(t^{*},p_i)},\ \;\;B=%
\frac{\partial ^2F}{\partial v^1\partial t}|_{(t^{*},p_i)},\ \;\;C=\frac{%
\partial ^2F}{\partial t^2}|_{(t^{*},p_i)}. 
\end{equation}
Then, taking notice of (\ref{100}), (\ref{102}) and (\ref{103}), we can
obtain the Taylor expansion of $F(v^1,t,\sigma )$ in the neighborhood of the
bifurcation point $(t^{*},p_i)$ 
\[
F(v^1,t,\sigma )=\frac 12A(v^1-p_i^1)^2+B(v^1-p_i^1)(t-t^{*})+\frac
12C(t-t^{*})^2 
\]
which by (\ref{99}) is the behavior of $\phi ^2$ in this region. Because of
the second equation of (\ref{88}), we get 
\[
A(v^1-p_i^1)^2+2B(v^1-p_i^1)(t-t^{*})+C(t-t^{*})^2=0 
\]
which leads to 
\begin{equation}
\label{104}A(\frac{dv^1}{dt})^2+2B\frac{dv^1}{dt}+C=0 
\end{equation}
and 
\begin{equation}
\label{105}C(\frac{dt}{dv^1})^2+2B\frac{dt}{dv^1}+A=0. 
\end{equation}
The different directions of the branch curves at the bifurcation point are
determined by (\ref{104}) or (\ref{105}). The remainder component $dv^2/dt$ can be given by
\[
\frac{dv^2}{dt}=f_1^2\frac{dv^1}{dt}+f_t^2 
\]
where partial derivative coefficients $f_1^2$ and $f_t^2$ have been
calculated in (\ref{98}).

In summary, that in our linear defect theory
there exist the crucial case of branch process. This means that, when an
original defect moves through the bifurcation point in the early universe,
it may split into two defects moving along different branch curves. Since
the topological current of linear defects is identically conserved, the sum
of the topological quantum numbers of these two splitted defects must be
equal to that of the original defect at the bifurcation point, i.e. 
\[
\beta _{i_1}\eta _{i_1}+\beta _{i_2}\eta _{i_2}=\beta _i\eta _i 
\]
for fixed $i$. This can be looked upon as the topological reason of linear
defect splitting. In the
bifurcation theory of topological current, there is another branch process
at a higher degenerated point. However, in the present case, since $\vec
\phi $ is a 2-dimensional vector field, there is no higher degenerated point
for linear defect current.

\end{document}